# Actionable Guidance for High-Consequence AI Risk Management: Towards Standards Addressing AI Catastrophic Risks

Anthony M. Barrett[†]*, Dan Hendrycks[†], Jessica Newman[†], Brandie Nonnecke[†]
[†] UC Berkeley
* Corresponding author: anthony.barrett@berkeley.edu

*This version last revised 23 February 2023*

## Abstract

  Artificial intelligence (AI) systems can provide many beneficial capabilities but also risks of adverse events. Some AI systems could present risks of events with very high or catastrophic consequences at societal scale. The US National Institute of Standards and Technology (NIST) has been developing the NIST Artificial Intelligence Risk Management Framework (AI RMF) as voluntary guidance on AI risk assessment and management for AI developers and others. For addressing risks of events with catastrophic consequences, NIST indicated a need to translate from high level principles to actionable risk management guidance.

  In this document, we provide detailed actionable-guidance recommendations focused on identifying and managing risks of events with very high or catastrophic consequences, intended as a risk management practices resource for NIST for AI RMF version 1.0 (released in January 2023), or for AI RMF users, or for other AI risk management guidance and standards as appropriate. We also provide our methodology for our recommendations.

  We provide actionable-guidance recommendations for AI RMF 1.0 on: identifying risks from potential unintended uses and misuses of AI systems; including catastrophic-risk factors within

---

the scope of risk assessments and impact assessments; identifying and mitigating human rights harms; and reporting information on AI risk factors including catastrophic-risk factors.

In addition, we provide recommendations on additional issues for a roadmap for later versions of the AI RMF or supplementary publications. These include: providing an AI RMF Profile with supplementary guidance for cutting-edge increasingly multi-purpose or general-purpose AI.

We aim for this work to be a concrete risk-management practices contribution, and to stimulate constructive dialogue on how to address catastrophic risks and associated issues in AI standards.

Note for readers referred to this document as an informative reference **for impact identification and characterization activities**, e.g., under AI RMF Map 1.1 or Map 5.1:

- **For identifying potential impacts,** see Section 3.2.2.1.1 of this document for impact dimensions to consider.
- **For a scale for assessing or rating the magnitude of identified impacts,** see Section 3.2.2.1.2 of this document. This impact magnitude scale closely follows the impact-rating Table H-3 of NIST SP 800-30, except for our addition of factors that could lead to severe or catastrophic consequences for society.

# Contents

# 1. Introduction and Objectives

The US National Institute of Standards and Technology (NIST) has been developing the NIST Artificial Intelligence Risk Management Framework, or AI RMF. NIST intends the AI RMF as voluntary guidance on AI risk assessment and other AI risk management processes for AI developers, deployers, users, and evaluators. NIST released Version 1.0 of the main AI RMF document in January 2023, and is periodically updating the more detailed guidance in the accompanying AI RMF Playbook. (NIST n.d.a, 2023a, 2023b)

As voluntary guidance, NIST does not impose "hard law" mandatory requirements for AI developers or deployers to use the AI RMF. However, AI RMF guidance is part of "soft law" norms and best practices, which AI developers and deployers have incentives to follow as appropriate. For example, insurers or courts may expect AI developers and deployers to show reasonable usage of relevant NIST AI RMF guidance as part of due care when developing or deploying AI systems in high-stakes contexts, in much the same way that NIST Cybersecurity Framework guidance can be used as part of demonstrating due care for cybersecurity.[2] In addition, elements of soft-law guidance are sometimes adapted into hard-law regulations, e.g., by mandating that particular industry sectors comply with specific standards.[3]

The AI RMF "core functions", or broad categories of activities, apply as appropriate across AI system lifecycles: "Map" for identifying AI risks in context; "Measure" for rating AI trustworthiness characteristics; "Manage" for decisions on prioritizing, avoiding, mitigating, or accepting AI risks; and "Govern" for AI risk management process policies, roles, and responsibilities. NIST decomposes these high-level functions into categories and subcategories of activities and outcomes. In addition, NIST provides more-detailed guidance in a companion Playbook resource document. The AI RMF helpfully prompts consideration of systemic and societal-scale risks, in addition to risks to individuals and groups. However, to date, its treatment of large-scale and catastrophic risks has been more in terms of challenges for the AI RMF rather than in detailed guidance. (NIST 2021a, NIST 2022a, NIST 2022b, 2023a, 2023b)

In the wake of significant advances of AI systems such as BERT, CLIP, GPT-3, DALL-E 2, and PaLM, we believe it is vitally important to address the possibility of both transformative benefits and catastrophic risks of AI, including the increasingly multi-purpose or general-purpose AI that can serve as AI platforms underpinning many end-use applications. Such advanced AI systems often have qualitatively distinct properties compared to narrower AI systems, such as the potential to be applied to many sectors at once, and emergent properties that can provide unexpected beneficial capabilities but also unexpected risks of adverse events. These AI systems could present corresponding catastrophic risks to society, such as correlated robustness failures across multiple high-stakes application domains (Bommasani et al. 2021 pp. 115-116). In the longer term, as advanced AI systems continue to grow in capability, their potential to pose catastrophic risks could grow, such as the potential for various kinds of misuse, or because the difficulty of safely controlling such systems would increase along with their capability (see, e.g., Russell 2019). Moreover, future multi-purpose AI systems are not the only AI systems that could cause large-scale harms to individuals and society. For example, misuse of AI systems, such as in disinformation campaigns or cyberattacks on critical infrastructure, could also pose risks of events with high consequences; for related discussion see, e.g., Brundage et al. (2018).

---

[2] NIST Cybersecurity Framework guidance "Profiles and associated implementation plans can be leveraged as strong artifacts for demonstrating due care." (NIST 2021d)
[3] For discussion and references on soft and hard law for AI governance, see, e.g., Gutierrez et al. (2021).

There are also vital relationships between principles of fairness and protecting human rights, addressing risks to individuals and groups, and addressing large-scale systemic or catastrophic risks. Some types of risks to individuals or groups comprise large-scale or catastrophic risks via accumulation or correlation of risks across individuals. Alignment of future AI systems with human values should include appropriate protection of human rights, and consideration of populations vulnerable to disproportionate harms. Indeed, preventing catastrophe can be an important part of preventing unfair outcomes; often the effects of catastrophe fall disproportionately on disadvantaged people.

In their AI RMF Concept Paper (NIST 2021a) and in AI RMF workshop discussions, NIST has indicated openness to addressing catastrophic risks and related issues in the AI RMF in some way. However, they also have indicated a need to translate from high level principles to actionable AI RMF guidance for AI developers, users, and evaluators, and that addressing AI catastrophic risks presents challenges.[4] Moreover, NIST has limited resources for developing guidance.

In this work, we assumed that NIST's current resources may not allow them to write in-depth guidance specifically to address catastrophic-risk factors and related issues for the AI RMF 1.0. We address that potential gap with this work by taking a proactive approach to drafting elements of actionable guidance on catastrophic risks and related issues for the AI RMF. Our actionable-guidance recommendations aim to complement NIST's more general AI RMF procedures for risk assessment and mitigation to constructively address catastrophic-risk factors and related issues that otherwise NIST may not address in the AI RMF.

Our main objectives for this work were to:
1) Develop small, simple pieces of guidance on catastrophic-risk factors and related issues for which it seems immediately tractable to draft appropriate, actionable guidance for use by AI developers, users, and evaluators. NIST could then incorporate these pieces of guidance into the AI RMF 1.0, extending its general procedures for risk assessment and mitigation to constructively address some catastrophic-risk factors.
2) Develop recommendations on additional issues for NIST to include in an AI RMF roadmap for later versions of the AI RMF or in supplemental publications, on the grounds that the issues are critical topics but appropriate guidance development would take additional time.

---

[4] E.g., in the AI RMF Concept Paper (NIST 2021a, p. 2) as part of discussion of unique challenges in managing AI risks: "Tackling scenarios that can represent costly outcomes or catastrophic risks to society should consider: an emphasis on managing the aggregate risks from low probability, high consequence effects of AI systems, and the need to ensure the alignment of ever more powerful advanced AI systems." That paragraph concludes by stating that NIST aims for the AI RMF to be "actionable guidance that is broadly adoptable".

This document also could serve as a complementary AI risk management practices resource for AI RMF users, or for users of other AI risk management guidance and standards.

In Section 2 of this document, we outline our approach, including criteria for recommendations. We provide our resulting actionable-guidance recommendations for AI RMF 1.0 in the "Guidance" subsections of Section 3. We provide our recommendations for versions of the AI RMF or supplementary publications after AI RMF Version 1.0 in Sections 4 and 5.

We aim for this work to be a concrete risk-management practices contribution for NIST and users of the AI RMF, as well as for users of related AI risk management standards such as ISO/IEC 23894, complementing other available AI risk management guidance and standards. We also aim to stimulate constructive dialogue on how to appropriately address AI catastrophic risks and associated issues in AI standards and AI governance more broadly. We seek feedback on this document, with the aim of improving and building on it in revisions of this document and in follow-on work.

# 2. Approach

In this work, we take a proactive approach to drafting elements of actionable guidance[5] on catastrophic risks and related issues. We identify potential ideas among the submissions we and others have already provided to NIST (e.g., at NIST 2021b and CLTC 2022) as well as additional ideas that emerge from discussions with colleagues or review of relevant literature. We focus primarily on developing small, simple pieces of guidance on catastrophic-risk factors and related issues for which appropriate guidance development seems immediately tractable for the AI RMF 1.0; we also identify critical topics for which it seems appropriate guidance development would take more time, and which we recommend NIST include in a roadmap for development of the AI RMF after releasing AI RMF version 1.0.

In the following section, we outline criteria we use to distinguish between those two categories.

## 2.1 Criteria for Actionable Guidance for AI RMF 1.0 vs. for Roadmap for Later Versions or Supplementary Publications

We only recommend that NIST include draft guidance on an issue in the AI RMF 1.0 if draft guidance appears to meet each of the following criteria, which we expect are important for NIST and key AI RMF stakeholders.[6] If we are not able to draft guidance now on an issue that meets all of these criteria, we recommend that NIST include that issue in an AI RMF roadmap for later versions of the AI RMF or supplementary publications. We evaluate draft guidance against the following criteria as we develop drafts of this guidance. We also ask reviewers to provide feedback to us on draft guidance using the main criteria from the following.

- Actionable and clear
  - Understandable and easy to use, as verified by reviewers and user testing
  - Would not impose unreasonable cost burdens nor limitations on innovations for beneficial applications of AI technologies
- Incorporating specific best practices and citable resources
  - Based largely on concepts and sources likely to pass eventual review by key external stakeholders (e.g., National Academies)
- Enterprise Risk Management (ERM)-compatible
  - Compatible/interoperable with, or able to be incorporated into, ERM frameworks used by many businesses and agencies
- Lifecycle-compatible

---

[5] As an example of the level of detail we would like to aim for as "actionable guidance" for a specific industry context, see documentation from the Open Web Application Security Project (OWASP) on identifying and prioritizing abuse cases for web-application software development (OWASP 2021). However, we also recognize that providing great detail for one industry context may render guidance unusable for other contexts. In this document we aim for a balance of enough detail for actionability without too much detail as to become unusable for many contexts.

[6] Based partly on stakeholder submissions to NIST about the AI RMF. See, e.g., the submissions from industry, public-interest groups, and other key stakeholders at NIST (2021b).

- ○ Usable for key stages of an AI lifecycle[7], which can include design, development, test and evaluation, deployment, continuous monitoring (and new iterations), and decommissioning
- Compatible with standards and regulations[8]
  - ○ Based on comparison to guidance from:
    - NIST
      - Including NIST Cybersecurity Framework, NIST Privacy Framework, NIST SP 800-30 and 800-53 on risk assessment, and NIST IR 8026 on privacy impact assessment
    - Institute of Electrical and Electronics Engineering (IEEE)
      - Including IEEE 7010 on AI / intelligent system impact assessment
    - International Standards Organization (ISO) and International Electrotechnical Commission (IEC)
      - Including ISO 31000 on risk management[9], ISO 26000 on social responsibility and human rights, ISO/IEC 23894 draft standard on AI risk management, and ISO/IEC 42001 draft standard on AI management system
    - EU AI Act
      - E.g., to consider "known and foreseeable risks", "conditions of reasonably foreseeable misuse", and "intended purpose" of high-risk AI systems, per draft EU AI Act language (EU 2021a)
    - European Committee for Standardization (CEN) and the European Committee for Electrotechnical Standardization (CENELEC)
      - CEN-CENELEC currently appears not to have other directly related standards of its own, but is engaged on standards development related to ISO/IEC and the EU AI Act
    - United Nations (UN) and the UN Educational, Scientific and Cultural Organization (UNESCO)
      - Including the UNESCO Recommendation on the Ethics of Artificial Intelligence (UNESCO 2021)
- Measurable or at least documentable
  - ○ The AI RMF Concept Paper (NIST 2021a) seems to provide a rough definition of measurability, i.e. something that can "indicate AI system trustworthiness in meaningful, actionable, and testable ways." Presumably that means something that can be checked or verified, e.g., by a programmer's automated code test or by an auditor reviewing relevant information, but not necessarily quantitative.
- ○ Where possible, we aim to provide steps that AI RMF users could employ to directly measure aspects of AI trustworthiness attributes or characteristics, and ideally also to measure risks (e.g., in terms of estimated probabilities and consequences of adverse events).
- ○ Otherwise, we aim to at least enable AI RMF users to document their use of a specific practice related to AI trustworthiness attributes or characteristics in the NIST AI RMF, in a way that can be checked or verified. (In other words, we aim to provide measurable processes for identifying and mitigating risks, even if the probability and consequences of those risks cannot be measured yet.[10])
- ● Small, tractable
  - ○ We can perform initial drafting of guidance, as well as revisions after receiving feedback, with little input from colleagues
  - ○ Review/testing by colleagues would not require long periods of time
- ● Modular, separable
  - ○ If one idea or one part of an idea does not prove viable, that would not render other parts or ideas inviable
- ● Low downside of including in AI RMF 1.0
  - ○ Future-proof enough to be used appropriately over the next 10 years[11], etc. such that guidance we write now is unlikely to be used or misused by stakeholders in ways that would be net-harmful
- ● Help NIST address legislative mandate
  - ○ Help NIST address parts of their mandate[12] related to identifying and providing procedures and processes for assessing trustworthiness (especially as related to characterizing safety, security, robustness, fairness and/or bias) and mitigating risks.

---

[7] For example, NIST (2021a) mentions the following major AI lifecycle stages: pre-design; design and development; test and evaluation; and deployment.

[8] NIST (2022b, p. 4) states that AI RMF guidance should be "law- and regulation-agnostic. The Framework should support organizations' abilities to operate under applicable domestic and international legal or regulatory regimes."

[9] On "risk" terminology: In ISO 31000 and related standards, "risk" can refer to potential for either a negative or positive deviation from the objectives or the expected. In this document we generally use the term "risk" more narrowly, as "risk of events with harms or adverse impacts/consequences", i.e., potential for negative deviation from the expected or the objectives. This usage of "risk" is generally consistent with usage in NIST publications or contexts with a focus on safety concerns, and does not preclude a broader usage of "risk" per ISO 31000.

## 2.2 Peer Review and User Testing

As part of a review and testing process for our draft guidance and recommendations, we have privately circulated initial draft material for feedback and user testing, asking reviewers and test users to check whether draft guidance meets all criteria for inclusion in AI RMF version 1.0. We continue to request feedback from organizations or individuals with expertise on related topics, including AI research and development, AI safety and security, risk assessment, human rights, governance, and catastrophic risks. (This includes posting a publicly available paper, with a goal of receiving additional feedback on our actionable-guidance recommendations and improving future versions of our recommendations.)

---

[10] The U.S. Chamber of Commerce's Technology Engagement Center ("C_TEC") submission in response to the NIST AI RMF RFI included a comment that "The RMF should specifically address situations where risk cannot be measured and offer guidance on reasonable steps for mitigating that risk without limiting innovation and investments in new and potentially beneficial AI technologies." (NIST 2021b, Comment 101)

[11] The 10 year timeframe assumes an approximately 10 year cycle for updates to AI RMF guidance and corresponding usage of updated guidance by AI RMF users. For comparison, NIST CSF updates have been every four years: NIST released CSF version 1.0 in 2014, version 1.1 in 2018, and is reportedly planning updates in 2022. ISO standards typically undergo update reviews every five years.

[12] i.e., touches on HR 6395, Division E, Section 5301(c) parts 1, 1B, 1C, and 2.

We invite readers of this document to provide us feedback on our criteria, the extent to which our guidance in this document meets our criteria, or how to improve this document.

# 3. Resulting Actionable-Guidance Recommendations for AI RMF 1.0

In our preliminary analysis, the following elements of guidance appear to meet many key criteria for actionable guidance for AI RMF 1.0 (small/tractable, modular/separable, incorporating specific best practices and citable resources, and helping NIST address legislative mandate). As we write draft guidance, we also keep other criteria in mind (i.e., ERM-compatible, lifecycle-compatible, compatible with standards and regulations, measurable or at least documentable, low downside of including in AI RMF 1.0, and actionable and clear).

We assumed that the following draft guidance would complement NIST's more general AI RMF procedures for risk assessment, mitigation, and organizational governance across an AI system lifecycle, that would be roughly analogous to procedures in the NIST Cybersecurity Framework (NIST 2018) and NIST Privacy Framework (NIST 2020a). Thus, while we provide some draft guidance on risk management and governance in this document as pertinent to catastrophic risks and related issues, in this document we do not provide more broadly-applicable guidance that we assumed would be redundant to broadly-applicable NIST AI RMF guidance in the AI RMF.

We present the following elements of guidance in roughly chronological order for when we expect users would first need to use them during an AI system development lifecycle. First is Section 3.1 for identification of potential unintended use cases and misuse cases, beyond an AI system designer's originally intended use case(s). Second is Section 3.2 for identification of risks and potential impacts, and rating magnitude of potential impacts, while considering societal-level catastrophic-risk factors. Third is Section 3.3 for human rights guidance, including factors to consider in identification and mitigation of human rights harms. Fourth is Section 3.4 for risk and incident reporting, including risk factors to consider reporting to users, incident databases or other stakeholders.

## 3.1 Identifying Potential Unintended Uses and Misuses of AI Systems

### 3.1.1 Overview and Rationale

This element of guidance aims to provide organizations with processes to anticipate potential unintended uses and misuses of an AI system. As part of this section, we provide guidance encouraging the early, proactive identification of additional beneficial uses, and other potential uses, besides an originally envisioned use. Later steps of the guidance encourage decision making for each identified category of uses; that could include disallowing particular categories of use, or implementing various controls. (This is broadly what OpenAI did with GPT-3, and would be compatible with the draft EU AI Act risk approach.) We also adapt relevant practices such as the cybersecurity concept of identifying and assessing risks of "misuse cases" or "abuse cases", namely ways that either an adversary or authorized user could maliciously or accidentally misuse an information system[13], in addition to considering intended use cases for authorized users of an information system.[14] (Organizations can then incorporate identified misuse/abuse cases into their harms modeling, security threat modeling, risk assessments, or related risk management processes as appropriate, such as threat events in context of risk assessments per NIST SP 800-30.)[15] One aim with this section is useful guidance on identifying and assessing risks of AI, yielding risk management strategies with some robustness regarding future potential uses and misuses beyond the AI designers' originally intended uses. This also could help identify sectors that could be affected by correlated robustness failures or other systemic risks, which we address again in later sections.[16]

We adapt guidance from resources such as the following:

- Microsoft provides some guidance on identifying potential types of harm, e.g., from intended uses, unintended uses, system errors, or misuses, as part of AI harms modeling; see Microsoft (2021a).
- OpenAI has codified the general idea of identifying misuse/abuse cases in their own AI safety best practices; their best practice #2 is "think like an adversary": see OpenAI (2020a). OpenAI exemplified this approach in their 2019 announcement of GPT-2, which included several categories of potential misuse cases: see the "Policy Implications" section of OpenAI (2019a).
- OpenAI also provided AI system developers with guidance on disallowed/unacceptable use-case categories of GPT-3: see the use case guidelines of OpenAI (2020b).
- Case studies documented in Newman (2020) detail how institutions including Microsoft and OpenAI have tried to improve the inclusiveness of AI design, development, use, and evaluation, and also reduce and manage the risk of potential negative impacts. At Microsoft for example, the Responsible AI Program includes the AETHER Committee, the Office of Responsible AI, a Responsible AI Standard, and a Responsible AI Champs community. Microsoft researchers have also documented the role of checklists in AI ethics and worked on "harms modeling" designed to help researchers anticipate the potential for harm and identify gaps in products that could put people at risk (Madaio et al. 2020, Microsoft 2021a).

---

[13] In this document we generally treat "abuse case" as synonymous with "misuse case".

[14] We assume the intended use case(s) would be distinct from potential unintended uses/misuse. We also assume that an AI system developer or deployer would already have one or more intended use cases identified, as they head into the process of applying this particular draft guidance to the AI system.

[15] Note that stakeholders could be impacted that are not intended users of an AI system, and one of the main goals of impact assessment is identifying impacts to individuals, groups, organizations or society beyond the users of an AI system. See Section 3.2 for more.

[16] For developers of increasingly multi-purpose or general-purpose AI such as GPT-3, PaLM, etc., it may be particularly important to assess a wide range of potential uses and misuses. These AI systems can serve as AI platforms underpinning many end-use applications, with broad benefits as well as systemic risks such as correlated robustness failures across multiple high-stakes application domains (see, e.g., Bommasani et al. 2021.)

- Available frameworks or taxonomies of AI-related adversary attack modes or other machine learning failure modes include MITRE ATLAS (MITRE 2021a) and Microsoft (2021c).
  - Updated versions of the NIST draft taxonomy of adversarial machine learning (Tabassi et al. 2019) or the NIST draft taxonomy of AI risk (NIST 2021a) also could be useful in defining misuse/abuse cases to consider for machine learning AI systems.
  - The Open Web Application Security Project (OWASP) provides guidance on identifying and prioritizing abuse cases for web-application software development (OWASP 2021).
  - Draft language for the EU AI Act also includes the general idea of considering "reasonably foreseeable misuse" along with an "intended purpose" of an AI system[17] (EU 2021a), though it does not yet provide detailed how-to guidance.

We also aim to enable risk assessment of an underlying AI system without having certainty about all specific use cases of AI systems incorporating an underlying AI system. In addition, we aim to ensure that this guidance could enable identification of important risks of increasingly general-purpose AI systems (including "foundation models"), which may have a wide range of potential application areas, at an early-enough AI lifecycle stage that risk mitigation would be more feasible and cost-effective than if waiting until specific intended use cases are marketed or deployed.

## 3.1.2 Guidance

Definitions for this section:
- "Intended user" includes any member of your organization, customers, or others generally using your system in ways and contexts that your organization intends for your system to be used.
- "Adversary" includes external attackers with malicious intent; it can also include unintended users more broadly, and intended users, if taking harmful actions with your system.
- "Misuse case" and "abuse case" have the same meaning in this document: a way that either an adversary or intended user could misuse an information system, either intentionally and maliciously, or unintentionally and accidentally.

Our guidance in this section assumes your team will consider possibilities of both malicious and accidental misuse when identifying misuse/abuse cases.[18] If your team finds it more appropriate

---

[17] The draft EU AI Act defines the terms as follows: "'intended purpose' means the use for which an AI system is intended by the provider, including the specific context and conditions of use, as specified in the information supplied by the provider in the instructions for use, promotional or sales materials and statements, as well as in the technical documentation", and "'reasonably foreseeable misuse' means the use of an AI system in a way that is not in accordance with its intended purpose, but which may result from reasonably foreseeable human behaviour or interaction with other systems". (EU 2021a)

[18] Although some sources and procedures define these terms differently and more narrowly, e.g., to focus on an actor's intent, others such as OWASP (2021) treat "abuse case" and "misuse case" as essentially equivalent. This approach serves purposes of efficiently identifying vulnerabilities and mitigations to both intentional and unintentional misuse as part of software development and testing. As an example of a

to consider accidental interactions separately from malicious misuse, then use those terms or other terms instead of the above, as long as you consistently consider possibilities of both malicious misuse and accidental interactions.

## 3.1.2.1 "Map" Guidance for Identifying Potential Unintended Uses and Misuses of AI Systems

**In addition to identifying an intended use case of an AI system, also:**

- **Identify other potentially beneficial use cases or applications of an AI system,** if your organization has not already done that (e.g., as part of assessing business opportunities) in a reasonably systematic manner.[19]
  - Use appropriate methods in order to:
    - Identify other potentially beneficial uses or applications that your organization might want to intentionally pursue
    - Identify other potentially beneficial uses or applications that your intended users might attempt, which your organization might not intentionally pursue
  - For methods to identify other potentially beneficial uses or use cases, consider methods such as:
    - Brainstorming
    - Reviewing publications that discuss current and potential uses of other AI systems. This can include industry publications and references such as ISO/IEC TR 24030.
    - Reviewing available information on competitors' uses of AI systems with similar characteristics
- **Identify misuse/abuse cases and types of adversary attack of an AI system**, using threat modeling, red team methods or other procedures as appropriate.[20]
  - Identify applicable misuse/abuse case types as listed in the following resources or in other relevant resources as appropriate:

---

software development and information security best practice that efficiently provides a layer of protection against both malicious misuse and accidental interactions, input validation ensures a software module only accepts inputs from other software modules or from users that meet specific criteria.

[19] This could be a risk management practice or control for the AI RMF Map 1.1 subcategory "Intended purposes, potentially beneficial uses, context-specific laws, norms and expectations, and prospective settings in which the AI system will be deployed are understood and documented. Considerations include: the specific set or types of users along with their expectations; potential positive and negative impacts of system uses to individuals, communities, organizations, society, and the planet; assumptions and related limitations about AI system purposes, uses, and risks across the development", or for the AI RMF Map 5.1 subcategory. (NIST 2023a)

[20] This could be a risk management practice or control for the AI RMF Map 1.1 subcategory "Intended purposes, potentially beneficial uses, context-specific laws, norms and expectations, and prospective settings in which the AI system will be deployed are understood and documented. Considerations include: the specific set or types of users along with their expectations; potential positive and negative impacts of system uses to individuals, communities, organizations, society, and the planet; assumptions and related limitations about AI system purposes, uses, and risks across the development", or for the AI RMF Map 5.1 subcategory. (NIST 2023a)

- MITRE ATLAS (MITRE 2021a) for AI-related adversary tactics and techniques
- Microsoft (2021c) for machine learning-related failure modes and threat taxonomy, and Microsoft (2021d, 2022a) for additional considerations in AI system threat identification and threat modeling
- MITRE ATT&CK® (MITRE 2021b) for general cybersecurity adversary tactics and techniques
- NIST SP 800-30 Appendix E for general information security threat events for both adversarial tactics, techniques and procedures as well as non-adversarial (i.e. accidental) threat events
- For illustrative examples of AI misuse cases, see, e.g., the "Policy Implications" section of the OpenAI (2019a) announcement of GPT-2, the "Case Studies" section of the MITRE ATLAS (MITRE 2021a) home page, or Brundage et al. (2018).
- Provide mechanisms such as email, web forms, or other hotlines for internal and external stakeholders to report concerns about potential types of AI misuse/abuse, or to report incidents of misuse/abuse, vulnerabilities discovered, etc. along with appropriate protections for stakeholders making reports. For example, see the vulnerability reporting procedures and safe-harbor policy of OpenAI (2020c).

- **For each potential misuse case, consider whether the misuse case could be a way for an adversary to attack your AI system, or a way for an adversary to use your AI system to attack something/someone else.**
- Compile a list of all misuse cases you identify, along with their key characteristics, such as whether the misuse case could be a way for an adversary to attack your AI system or a way for an adversary to use your AI system to attack something/someone else.
- For more detailed example procedures on how to identify and analyze abuse cases, see OWASP guidance on identifying and prioritizing abuse cases for web-application software development (OWASP 2021). However, keep in mind that some adaptations of OWASP guidance will likely be appropriate for analyzing AI applications. For example, instead of basing abuse cases on the OWASP Top 10 web application security risk types, consider basing abuse cases on AI-relevant security risk types such as listed in MITRE ATLAS (MITRE 2021a).

**When to identify use cases and misuse/abuse cases of an AI system**:[21]

---

[21] This could be a risk management practice or control for the AI RMF Map 1.1 subcategory "Intended purposes, potentially beneficial uses, context-specific laws, norms and expectations, and prospective settings in which the AI system will be deployed are understood and documented. Considerations include: the specific set or types of users along with their expectations; potential positive and negative impacts of system uses to individuals, communities, organizations, society, and the planet; assumptions and related limitations about AI system purposes, uses, and risks across the development", or for the AI RMF Map 5.1 subcategory. (NIST 2023a)

- Identify potential use cases during early stages of your AI system lifecycle, such as plan and design, at minimum.
- Identify misuse cases during all major stages of your AI system lifecycle (or approximate equivalents in Agile/iterative development sprints), such as: plan, data collection, design, train/build/buy, test and evaluation, deploy, operate and monitor, and decommission.
- Plan to revisit use and misuse case identification at key intended milestones, or at periodic intervals (e.g., at least annually), whichever comes first.

**For staffing to identify potential use cases and misuse cases of an AI system**:[22]
- Include members of each of the following functional teams (or equivalents) as appropriate:
  - For identifying *both* **potentially beneficial use cases** *and* **for identifying misuse cases** to consider in risk assessment or other assessments:
    - Product development, operations, human-computer interaction, user experience, policy, and ethics professionals
  - In addition, **for identifying misuse cases** to consider in risk assessment or other assessments:
    - Security
  - In addition, **for identifying other potentially beneficial use cases** to consider in risk assessment or other assessments:
    - Marketing and sales
- Consider including members of other teams as appropriate, such as:
  - Research and development (for additional technically-informed perspective on AI system capabilities and limitations)
  - External-facing teams and/or external stakeholders (for additional early identification of potential stakeholder concerns and other stakeholder perspectives)

**After identifying use cases and misuse cases of an AI system:**
- **For each identified potentially beneficial use case of an AI system:**
  - **If it seems plausible that your organization or your intended users might pursue/attempt those uses, consider carrying out impact assessments, risk assessments, or other assessments of those use cases, or incorporating these new uses into your overall risk assessment processes, as appropriate.** For example, if there is significant potential for using an AI system in a more sensitive application than originally intended, that could modify the overall assessment of risk of the AI system in question, even if its originally intended use case seems low risk.[23]

---

[22] This could be a risk management practice or control for the AI RMF Map 1.1 subcategory "Intended purposes, potentially beneficial uses, context-specific laws, norms and expectations, and prospective settings in which the AI system will be deployed are understood and documented. Considerations include: the specific set or types of users along with their expectations; potential positive and negative impacts of system uses to individuals, communities, organizations, society, and the planet; assumptions and related limitations about AI system purposes, uses, and risks across the development", or for the AI RMF Map 5.1 subcategory. (NIST 2023a)

- E.g., consider characterizing each use case according to the OECD framework for the classification of AI systems (OECD 2022)

- **For each identified misuse/abuse case of an AI system:**
  - **Incorporate identified misuse/abuse cases and their key characteristics into your organization's harms modeling, security threat modeling, risk assessments, risk register, or related risk management processes as appropriate** (e.g., as threat events in context of risk assessments per NIST SP 800-30).[24]
    - As part of AI system threat modeling and risk assessment, consider resources such as Microsoft (2022a) and Microsoft (2021d).

### 3.1.2.2 "Manage" Guidance for Identifying Potential Unintended Uses and Misuses of AI Systems

After identifying and analyzing use cases and misuse cases of an AI system (per "Map" function guidance) :

- **For each identified potentially beneficial use case of an AI system:**
  - **Consider defining and communicating to key stakeholders whether any potential use cases (or categories of use cases) would be unacceptable, or would be treated as "high risk" or another category for which your organization would provide specific risk management guidance or other risk mitigation measures.**[25]
    - E.g., consider whether any potential uses would be regarded under the EU AI Act as falling into one of the following risk categories: "unacceptable risk", "high risk", or "low or minimal risk", per draft EU AI Act language (EU 2021a Section 5.2.2). For example, AI systems would

---

[23] This could be a risk management practice or control for the AI RMF Map 1.1 subcategory "Intended purposes, potentially beneficial uses, context-specific laws, norms and expectations, and prospective settings in which the AI system will be deployed are understood and documented. Considerations include: the specific set or types of users along with their expectations; potential positive and negative impacts of system uses to individuals, communities, organizations, society, and the planet; assumptions and related limitations about AI system purposes, uses, and risks across the development", or for the AI RMF Map 5.1 subcategory. (NIST 2023a)

[24] This could be a risk management practice or control for the AI RMF AI RMF Map 1.1 subcategory "Intended purposes, potentially beneficial uses, context-specific laws, norms and expectations, and prospective settings in which the AI system will be deployed are understood and documented. Considerations include: the specific set or types of users along with their expectations; potential positive and negative impacts of system uses to individuals, communities, organizations, society, and the planet; assumptions and related limitations about AI system purposes, uses, and risks across the development", or for the AI RMF Map 5.1 subcategory. (NIST 2023a) Alternatively, this could be a risk management practice or control for the AI RMF Measure 2.7 subcategory "AI system security and resilience – as identified in the Map function – are evaluated and documented." (NIST 2023a)

[25] This could be a risk management practice or control for the AI RMF Manage 1.3 subcategory "Responses to the AI risks deemed high priority, as identified by the Map function, are developed, planned, and documented. Risk response options can include mitigating, transferring, avoiding, or accepting." (NIST 2023a)

fall in the unacceptable-risk category if their use would violate fundamental rights.

## 3.2 Including Catastrophic-Risk Factors in Scope of Risk Assessments and Impact Assessments

### 3.2.1 Overview and Rationale

This element of guidance aims to complement an organization's previous processes for risk identification, assessment, and prioritization related to AI systems (or analogous processes as part of impact assessment[26]), to help ensure that the scope and time frame considered in those processes address reasonably foreseeable risks that could have high impacts or consequences for society. Such risks could include novel or extreme "Black Swan" events, or systemic risks, or longer-term impacts, even if such risks may seem to AI designers to be outside the typical scope of consideration for their organization (e.g., if they typically focus on common events, or on consequences to the organization). Including such risks in the scope of risk assessments and impact assessments could provide risk-management benefits to organizations, because risks with large impacts to society could also lead to large indirect impacts to the organization (e.g., by negatively affecting the organization's reputation and sales, or liabilities).

See Appendix 1 for notes on approaches to risk assessment scope and time frame in several relevant risk management publications, which we expect would be part of guidance for an organization's previous processes for risk identification, assessment, and prioritization related to AI systems. We adapt and build upon that guidance in the following.[27]

We aim our guidance in this section at organizations that plan to perform substantial risk assessments and impact assessments using industry-standard risk management guidance such as listed in Appendix 1, and organizations developing or deploying potentially high-risk AI systems such as cutting-edge AI systems that could have many high-impact uses or misuses beyond an originally intended use case. Some organizations may decide to use a prioritization or triage process to decide how in-depth a risk assessment or impact assessment should be for a specific AI system, to avoid excessive costs of in-depth assessments for AI systems that seem likely to present low risks.[28] Organizations may find it useful to distill and adapt our guidance in this section as part of a set of triage questions to identify potentially high-risk AI systems to prioritize for in-depth risk assessment or impact assessment.[29]

---

[26] Some authors use the term "AI impact assessment" specifically in context of *ex post* assessment of real-world consequences or effects of realized AI system risks, as determined after deployment of an AI system. Our recommendations for impact assessment scope could be applied in an *ex post* context. However, in this document we generally use the terms "potential impact" or "impact assessment" to refer to assessment of potential adverse consequences or potential harms of AI system risks, in context of an *ex ante* risk assessment or *ex ante* impact assessment performed before deployment and use of an AI system leads to real-world occurrence of those harms. Our usage of "potential impact" in context of *ex ante* risk assessment or *ex ante* impact assessment is broadly consistent with NIST SP 800-30 and related NIST publications on information system risk assessment, with NIST SP 1270 discussion of algorithmic impact assessment, and with ISO/IEC 23894 on AI risk management.

[27] E.g., we draw the phrase "Damage to or incapacitation of a critical infrastructure sector" from Table H-2 of NIST SP 800-30, and the phrase "Impacts on democratic institutions and quality of life" from NISTIR 8062.

## 3.2.2 Guidance

### 3.2.2.1 "Map" Guidance for Including Catastrophic-Risk Factors in Scope of Risk Assessments and Impact Assessments

#### 3.2.2.1.1 When Identifying Risks for an AI System Development or Deployment Project or Enterprise

**Regarding scope of AI system potential-impact assessments or risk assessments:**

- **Identify or assess reasonably foreseeable potential impacts to individuals, groups, organizations and society, as appropriate.**
  - Identify or assess reasonably foreseeable potential adverse impacts or harms of the following types:[30]
    - To organizational operations, including:
      - Missions and functions
      - Image and reputation, including:
        - Loss of trust and reluctance to use the system or service
        - Internal culture costs that impact morale or productivity
    - To organizational assets, including legal compliance costs arising from problems created for individuals
    - To other organizations
    - To individuals, including impacts to health, safety, well-being, or fundamental rights
    - To groups, including populations vulnerable to disproportionate adverse impacts or harms
    - To the Nation or other societal impacts, including:

---

[28] BSA | The Software Alliance's comment to NIST on the AI RMF Initial Draft included a recommendation "that NIST provide high level guidance to help organizations identify the 'consequential' AI systems that should be prioritized for assessment under the RMF. Such guidance could highlight suggested criteria and/or questions for assessing whether a system is consequential. For instance: …. Could the system pose a risk of significant physical or psychological injury or otherwise threaten an individual's human rights?" (Troncoso 2022)

[29] Additional resources for criteria to identify high-risk AI systems or use cases include the OECD AI risk criteria, with which our guidance is broadly consistent: scale or seriousness of adverse impacts; scope or number of individuals affected; and optionality of whether to be affected (OECD 2022, p. 67).

[30] This could be a risk management practice or control for the AI RMF Map 1.1 subcategory "Intended purposes, potentially beneficial uses, context-specific laws, norms and expectations, and prospective settings in which the AI system will be deployed are understood and documented. Considerations include: the specific set or types of users along with their expectations; potential positive and negative impacts of system uses to individuals, communities, organizations, society, and the planet; assumptions and related limitations about AI system purposes, uses, and risks across the development", or for the AI RMF Map 5.1 subcategory. (NIST 2023a)

- Damage to or incapacitation of a critical infrastructure sector
- Economic and national security
- Impacts on democratic institutions and quality of life
- Environmental impacts

■ If appropriate for your context, also identify or assess more specific types of harms within the above categories (e.g., per NIST SP 800-30 Table H-2) or other types of harms (e.g., as outlined in Microsoft 2021a, 2021b, or in other resources in Appendix III of PAI 2022)

**Regarding criteria or objectives for scenario identification and screening/prioritization:**

- **Aim to identify scenarios with high consequences for society, or with factors that could lead to high consequences.**[31]
    - **For more on factors that could lead to high consequences for society, see factors under the "High" impact category description in the next subheading of this document** ("When assessing or rating the magnitude of potential impacts of an AI system development or deployment project or enterprise")
    - Do not only focus on what seem like the most likely scenarios; do not ignore scenarios that would be novel or extreme "Black Swan" events
    - Do not ignore scenarios for which you cannot measure the magnitude of the consequences or probability yet, especially if the impacts could be severe or catastrophic[32]

**Regarding potential-impact or risk assessment time frames and when to reassess risks:**

- At minimum: Assess risks for each expected stage of your AI system lifecycle (or approximate equivalents in Agile/iterative development sprints), such as: plan, design, train/build/buy, test and evaluation, deploy, operate and monitor, and decommission; and plan to revisit impact/risk assessment at key intended milestones, or at periodic intervals (e.g., at least annually), whichever comes first[33]

---

[31] This could be a risk management practice or control for the AI RMF Map 1.1 subcategory "Intended purposes, potentially beneficial uses, context-specific laws, norms and expectations, and prospective settings in which the AI system will be deployed are understood and documented. Considerations include: the specific set or types of users along with their expectations; potential positive and negative impacts of system uses to individuals, communities, organizations, society, and the planet; assumptions and related limitations about AI system purposes, uses, and risks across the development", or for the AI RMF Map 5.1 subcategory. (NIST 2023a)

[32] This also could be part of risk management practices or controls for the AI RMF Measure 3.2 subcategory "Risk tracking approaches are considered for settings where AI risks are difficult to assess using currently available measurement techniques or where metrics are not yet available." (NIST 2023a) This seems potentially valuable as part of a proactive effort to manage risks that are not yet easily assessed. It also seems likely actionable for many key AI development organizations employing enterprise risk management, e.g., by tracking identified risks using a risk register. (For more on risk registers, see, e.g., ISO Guide 73 Section 3.8.2.4, PMI 2017 p. 417, and Stine et al. 2020.)

[33] This could be a risk management practice or control for the AI RMF Map 1.1 subcategory "Intended purposes, potentially beneficial uses, context-specific laws, norms and expectations, and prospective

- **Consider additional steps to identify or assess longer term impacts or use longer time horizons, and to reduce potential for surprise, if these additional assessment steps would be valuable for risk management (e.g., in high-stakes contexts, or for very powerful or increasingly multi-purpose / general-purpose AI systems**)[34]
    - Consider whether any risk assessment or impact assessment answers would change if assessing longer-term time periods (e.g., beyond the next year)
        - What additional impacts would you expect if your AI system is deployed for a long period of time?
        - What impacts would you expect to have greater magnitude if your AI system is deployed for a long period of time?
    - Identify unintended potential future events that should trigger reassessment or other responses, and build them into risk registers and/or planning and implementation of relevant lifecycle stages. To identify trigger events, consider questions such as:
        - What if monitoring indicates one of your risk-mitigation controls is not working as expected? (Consider this, as applicable, for each relevant risk-mitigation control.)
        - What if AI capability developments occur that are not expected until further into the future, such as availability of much more powerful AI systems or cloud computing?

3.2.2.1.2 When Assessing or Rating the Magnitude of Potential Impacts of an AI System Development or Deployment Project or Enterprise

**For rating potential impacts in a way that includes criteria for rating potential risks as catastrophic if they could lead to high adverse consequences for society:**

- **Consider using the following impact assessment scale** (which closely follows Table H-3 of NIST SP 800-30 except for our addition of "Factors that could lead to severe or catastrophic consequences for society include" and associated factors)[35], or a similar

---

settings in which the AI system will be deployed are understood and documented. Considerations include: the specific set or types of users along with their expectations; potential positive and negative impacts of system uses to individuals, communities, organizations, society, and the planet; assumptions and related limitations about AI system purposes, uses, and risks across the development", or for the AI RMF Map 5.1 subcategory. (NIST 2023a) scale, or quantitative impact/consequence estimates (e.g., for health and safety risks) and impact category threshold values:[36]

[34] This could be a risk management practice or control for the AI RMF Map 1.1 subcategory "Intended purposes, potentially beneficial uses, context-specific laws, norms and expectations, and prospective settings in which the AI system will be deployed are understood and documented. Considerations include: the specific set or types of users along with their expectations; potential positive and negative impacts of system uses to individuals, communities, organizations, society, and the planet; assumptions and related limitations about AI system purposes, uses, and risks across the development", or for the AI RMF Map 5.1 subcategory. (NIST 2023a)

[35] Except for our additions on catastrophic-risk factors, our only change from the language in Table H-3 of NIST SP 800-30 is that we use "Nation or society as a whole" instead of "Nation"; this may better match the scope of the AI RMF's aim to assess societal impacts, and draws on "society as a whole" language in the AI RMF Initial Draft (NIST 2022a).

- **"Very high"** impact category description: The threat event could be expected to have **multiple severe or catastrophic** adverse effects to organizational operations, organizational assets, individuals, other organizations, or the Nation or society as a whole. See "high" impact category below for more on severe or catastrophic adverse effects.
- **"High"** impact category description: The threat event could be expected to have a **severe or catastrophic** adverse effect on organizational operations, organizational assets, individuals, other organizations, or the Nation or society as a whole.
  - A severe or catastrophic adverse effect means that, for example, the threat event might: (i) cause a severe degradation in or loss of mission capability to an extent and duration that the organization is not able to perform one or more of its primary functions; (ii) result in major damage to organizational assets; (iii) result in major financial loss; or (iv) result in severe or catastrophic harm to individuals involving loss of life or serious life-threatening injuries.
  - **Factors that could lead to severe or catastrophic consequences for society include:**
    - **Potential for correlated robustness failures or other systemic risks across high-stakes application domains** such as critical infrastructure or essential services[37]
    - **Potential for other systemic impacts, which can be accumulated, accrued, correlated or compounded at societal scale**[38], e.g.:

---

[36] This could be a risk management practice or control for the AI RMF Map 5.1 subcategory "Likelihood and magnitude of each identified impact (both potentially beneficial and harmful) based on expected use, past uses of AI systems in similar contexts, public incident reports, feedback from those external to the team that developed or deployed the AI system, or other data are identified and documented." (NIST 2023a)

[37] See, e.g., discussion of correlated failures of "foundation models" spanning multiple critical functions in Section 4.9.3 of Bommasani et al. (2021).

[38] This draws upon the following types of systemic risks to society as described by Mallah (2022) and FLI (2022):
- Accumulated risk: small harms accumulating over time to form a major harm;
- Accrued risk: where events that are low-probability in the short-term, but high-impact, can accrue and build to significant-probability in the medium term;
- Correlation risk: where there are adverse events that are not evident in unit tests or accuracy tests, but can be expected to emerge from correlated decisions or correlated actions with a large number of users, instances, or executions of the system;
- Latent risk: where harms that will not manifest significantly or at all on system training or release may still be expected to appear with distributional shift, new use cases, or qualitative shifts in capabilities arising from quantitative scaling;
- Compounding risk: where harms would be expected to manifest only when either other problems occur or unexpected, but conceivable conditions or interactions manifest.

Mallah (2022) and FLI (2022) also mentioned adversarial risk, where harms manifest due to the lack of robustness in the system when in the presence of optimization pressures for inputs to induce those

- ○ **Potential for correlated bias** across large numbers of people or a large fraction of a society's population[39]
  - ○ **Potential for many high-impact uses or misuses beyond an originally intended use case**, e.g., if an AI system is a cutting-edge large-scale language model, "foundation model" or another highly multi-purpose / general-purpose AI system[40], or if it enables recursive improvement of capabilities of cutting-edge AI system algorithms or architecture through code generation, architecture search, etc.[41]
  - ● **Potential for large harms from mis-specified goals** (e.g., using over-simplified or short-term metrics as proxies for desired longer-term outcomes)[42]
  - ○ **"Moderate"** impact category description: The threat event could be expected to have a **serious** adverse effect on organizational operations, organizational assets, individuals, other organizations, or the Nation or society as a whole. A serious adverse effect means that, for example, the threat event might: (i) cause a significant degradation in mission capability to an extent and duration that the organization is able to perform its primary functions, but the effectiveness of the functions is significantly reduced; (ii) result in significant damage to organizational assets; (iii) result in significant financial loss; or (iv) result in significant harm to individuals that does not involve loss of life or serious life-threatening injuries.[43]

---

harms. Adversarial-risk impacts can be rated using our rating scheme in this section if your organization is not already using an equivalent impact rating scheme for adversarial risks. In creating our impact rating scheme, we aimed for consistency with impact rating of cybersecurity risks, including adversarial risks, in NIST SP 800-30 (NIST 2012, Table H-3).

Note that these types of systemic risks are not necessarily exclusive of one another. For example, adversarial risks can also be correlation risks, latent risks, compounding risks, etc.

[39] E.g., as discussed by Schwartz et al. (2022, p. 32): "The systemic biases embedded in algorithmic models can … be exploited and used as a weapon at scale, causing catastrophic harm."

[40] We believe that most AI systems could be readily identified as being in one of the following categories:
  A. One of a few large-scale, cutting-edge, increasingly multi-purpose or general-purpose AI system platforms (including "foundation models"), such as BERT, CLIP, GPT-3, DALL-E 2, and PaLM.
  B. A relatively narrow-purpose end-use application that builds on a multi-purpose AI model platform.
  C. One of many small-scale and/or stand-alone narrow-purpose AI systems.

Category A presents substantial potential for systemic impacts to society (see, e.g., Bommasani et al. 2021).

[41] As the DeepMind paper on the software code-generation AI system AlphaCode stated, "Longer term, code generation could lead to advanced AI risks. Coding capabilities could lead to systems that can recursively write and improve themselves, rapidly leading to more and more advanced systems." (Li et al. 2022) For discussion of related issues, see, e.g., Russell (2019).

[42] For examples of mis-specified objectives, such as social-media content recommendation machine-learning algorithms that learn to optimize user-engagement metrics by serving users with extremist content or disinformation, see, e.g., Rudner and Toner (2021d). Identifying mis-specification risks can also be aided by considering the following questions for an AI system: "What objective has been specified for the system, and what kinds of perverse behavior could be incentivized by optimizing for that objective?" Rudner and Toner (2021d, p. 10)

[43] Table H-3 of NIST SP 800-30, which inspired the written impact assessment scale used here, qualifies "moderate" impact as threat events that could be expected to cause a *serious* or *significant* adverse

- ○ **"Low"** impact category description: The threat event could be expected to have a **limited** adverse effect on organizational operations, organizational assets, individuals, other organizations, or the Nation or society as a whole. A limited adverse effect means that, for example, the threat event might: (i) cause a degradation in mission capability to an extent and duration that the organization is able to perform its primary functions, but the effectiveness of the functions is noticeably reduced; (ii) result in minor damage to organizational assets; (iii) result in minor financial loss; or (iv) result in minor harm to individuals.
- ○ **"Very Low"** impact category description: The threat event could be expected to have a **negligible** adverse effect on organizational operations, organizational assets, individuals, other organizations, or the Nation or society as a whole.

### 3.2.2.2 "Govern" Guidance for Including Catastrophic-Risk Factors in Scope of Risk Assessments and Impact Assessments

On other aspects of AI system potential-impact/risk assessment process planning:

- For contexts where the organization will need to characterize an AI system according to another AI classification framework, such as the OECD framework (OECD 2022), use risk assessment outputs as part of preparation for AI classification reporting. (Or if the AI system is already classified with another framework, use the AI classification information to inform risk assessment.)[44]

# 3.3 Human Rights Guidance Operationalized

## 3.3.1 Overview and Rationale

With this element of guidance, we provide actionable guidance on protecting human rights in the context of AI systems. Protecting human rights is a critical component of mitigating societal and catastrophic risks and is a commonly used framework by many companies and other stakeholders, including through implementation of human rights impact assessments. In addition to addressing risks from currently available AI systems, this could help ensure alignment of future AI systems with human values and principles such as fairness. For example, definitions for several AI trustworthiness concepts such as "privacy", "fairness", and "bias" relate to recognized human rights in international law. Privacy is a protection granted under Article 12 of the Universal Declaration of Human Rights (UDHR), and fairness and bias are related to

---

effect. Although we have not modified the language in the "Moderate" impact category description here to avoid inconsistency with NIST SP 800-30, we recommend NIST consider removing or replacing the words "serious" and "significant" in Table H-3 of NIST SP 800-30 with other terms, to position the "moderate" category more evenly between the "low" and "high" impact categories.

[44] This could be a risk management practice or control for the AI RMF Govern 4.2 subcategory "Organizational teams document the risks and potential impacts of the AI technology they design, develop, deploy, evaluate, and use, and they communicate about the impacts more broadly", or for the Govern 4.3 subcategory "Organizational practices are in place to enable AI testing, identification of incidents, and information sharing." (NIST 2023a)

rights of "non-discrimination" protected under Article 2 of the UDHR and Article 26 in the International Covenant on Civil and Political Rights (ICCPR) (UN 1948; UN 1966). These definitions have been codified over decades through the human rights legal framework (e.g., in charters, national laws and regulations). Additionally, the UN Guiding Principles on Business and Human Rights (UNGPs), established in 2011, are particularly salient for commercial AI development as they provide guidance to corporations on how to appropriately respect human rights in their operations (UN 2011). By analyzing how responsible AI principles have been interpreted through the human rights legal framework, we can use definitions that have reached fairly widespread consensus over decades of negotiation. Relevant work on human rights in AI includes: Nonnecke and Dawson (2021), Latonero (2018), Donahoe and Metzger (2019), Mantelero and Esposito (2021), and Bradley et al. (2021). Examples of relevant guidance or applications of such guidance include BSR (2012, 2013, 2018), Götzmann et al. (2016), and Microsoft (2018).

## 3.3.2 Guidance

The following guidance is based heavily on the UDHR (UN 1948) and the UNGPs (UN 2011).[45] Basics of the UNGPs include:

- Avoid infringing on human rights and address human rights impacts with which your organization is involved (UNGP 11).
- In all contexts, business enterprises should: (a) Comply with all applicable laws and respect internationally recognized human rights, wherever they operate; (b) Seek ways to honor the principles of internationally recognized human rights when faced with conflicting requirements; (c) Treat the risk of causing or contributing to gross human rights abuses as a legal compliance issue wherever they operate. (UNGP 23)

### 3.3.2.1 "Map" Guidance for Human Rights Guidance Operationalized[46]

- **Conduct human rights due diligence** (e.g., human rights impact assessments) at appropriate stages throughout an AI system lifecycle, beginning as early as possible (UNGPs 17, 18, and 24).
- **Assess potential and actual impacts to human rights at regular intervals as appropriate** throughout an AI system lifecycle (e.g., prior to product launch and throughout implementation), as part of human rights due diligence (UNGPs 17 - 20).
- **Identify potential or actual human rights impacts** (UNGPs 17- 20). **Potential example questions and UDHR Articles to consider include:**
    - UDHR Article 2, including non-discrimination and equality before the law

---

[45] "The UN Guiding Principles on Business and Human Rights have emerged as the global standard for companies' management of their human rights impacts." (BSR 2013, p.5)

[46] This could be a risk management practice or control for the AI RMF Map 1.1 subcategory "Intended purposes, potentially beneficial uses, context-specific laws, norms and expectations, and prospective settings in which the AI system will be deployed are understood and documented. Considerations include: the specific set or types of users along with their expectations; potential positive and negative impacts of system uses to individuals, communities, organizations, society, and the planet; assumptions and related limitations about AI system purposes, uses, and risks across the development", or for the AI RMF Map 5.1 subcategory. (NIST 2023a)

- ■ How could an AI system's bias in data or unfair algorithmic decisions affect rights to equal protection and non-discrimination?
  - ○ UDHR Article 3, including right to life and personal security
    - ■ How could an AI system's algorithmic decisions affect the right to life and personal security?
  - ○ UDHR Article 12, including privacy and protection against unlawful governmental surveillance
    - ■ How could an AI system be used for surveillance, leading to loss of privacy or inadequate protection of personally identifiable information?
  - ○ UDHR Articles 18 and 19, including freedom of thought, conscience and religious belief and practice, and freedom of expression and to hold opinions without interference
    - ■ How could an AI system affect rights to express opinions or practice religion?
  - ○ UDHR Articles 20 and 21, including freedom of association and the right to peaceful assembly
    - ■ How could an AI system affect rights to association, peaceful assembly, and democratic participation in government?
  - ○ UDHR Articles 23 and 25, including rights to decent work and to an adequate standard of living
    - ■ How could an AI system affect rights to decent work, including effects on adequate standard of living via displacement of human workers?

### 3.3.2.2 "Manage" Guidance for Human Rights Guidance Operationalized

- ● After identifying potential human rights harms:
  - ○ **Take appropriate action to prevent, address, cease, or mitigate human rights harms as part of human rights due diligence** (UNGPs 17 - 20).[47]
  - ○ Track effectiveness of responses as part of human rights due diligence (UNGP 20) and as part of continuous monitoring of risks after AI system deployment.[48]
  - ○ Establish and provide access to grievance remedy procedures for human rights harms (UNGPs 28 - 31).[49]

---

[47] This could be a risk management practice or control for the AI RMF Manage 1.3 subcategory "Responses to the AI risks deemed high priority, as identified by the Map function, are developed, planned, and documented. Risk response options can include mitigating, transferring, avoiding, or accepting", or for the Manage 2.2 subcategory "Mechanisms are in place and applied to sustain the value of deployed AI systems." (NIST 2023a)

[48] This could be a risk management practice or control for the AI RMF Manage 4.1 subcategory "Post-deployment AI system monitoring plans are implemented, including mechanisms for capturing and evaluating input from users and other relevant AI actors, appeal and override, decommissioning, incident response, recovery, and change management", or for the Manage 2.2 subcategory "Mechanisms are in place and applied to sustain the value of deployed AI systems." (NIST 2023a)

[49] This could be a risk management practice or control for the AI RMF Manage 4.1 subcategory "Post-deployment AI system monitoring plans are implemented, including mechanisms for capturing and evaluating input from users and other relevant AI actors, appeal and override, decommissioning, incident response, recovery, and change management", or for the Manage 2.2 subcategory "Mechanisms are in place and applied to sustain the value of deployed AI systems." (NIST 2023a)

### 3.3.2.3 "Govern" Guidance for Human Rights Guidance Operationalized

- After identifying and managing potential human rights harms:
  - Communicate risks of severe human rights harms and how the organization addresses them as part of human rights due diligence (UNGP 21).[50]

# 3.4 Reporting Information on AI Risk Factors and Incidents

## 3.4.1 Overview and Rationale

This element of guidance aims to help in communication to internal and external stakeholders as appropriate regarding factors and circumstances with potential for high-consequence risks, even if it is only to say that the possible risk requires additional attention and research in their development of AI products and services.

- For risk factors, we incorporate material from our draft risk-assessment guidance in Section 3.2 of this document.
- For incident reporting, we reference information used in the Partnership on AI's AI Incident Database (AIID 2021a).

Our approach aims for compatibility with the draft EU AI Act language (EU 2021a, Ch.2, Articles 9 and 13), which requires reporting residual risks of high-risk AI systems to users.

## 3.4.2 Guidance

### 3.4.2.1 "Govern" Guidance for Reporting Information on AI Risk Factors and Incidents

**Report risk factors identified in AI system risk assessment, including on the following potential types of impacts or harms outside the organization, by time of deployment or at earlier lifecycle stages, as appropriate in context as part of communicating AI system limitations and risks to stakeholders**:[51]

- Report information related to the applicable AI system use case or use cases (e.g., as in the OECD AI Framework for the Classification of AI Systems) where relevant.
- Also report AI system risk-factor information for other identified potential use cases for identified misuse/abuse cases as a whole (without non-obvious details of misuse/abuse cases that could be useful to adversaries), and/or for underlying AI systems separate from use/misuse/abuse cases, as appropriate to provide stakeholders a more complete

---

[50] This could be a risk management practice or control for the AI RMF Govern 4.2 subcategory "Organizational teams document the risks and impacts of the AI technology they design, develop, deploy, evaluate, and use, and they communicate about the impacts more broadly", or for the Govern 4.3 subcategory "Organizational practices are in place to enable AI testing, identification of incidents, and information sharing." (NIST 2023a)

[51] This could be a risk management practice or control for the AI RMF Govern 4.2 subcategory "Organizational teams document the risks and impacts of the AI technology they design, develop, deploy, evaluate, and use, and they communicate about the impacts more broadly", or for the Govern 4.3 subcategory "Organizational practices are in place to enable AI testing, identification of incidents, and information sharing." (NIST 2023a)

picture of reasonably foreseeable risks without providing adversaries too much information.

- ○ To stimulate consideration of what details would provide adversaries with too much information, see resources such as the Questions for Thinking About Risk subsection of the Resources section of the Partnership on AI's Publication Norms for Responsible AI (PAI 2022).
- For organization-internal stakeholders or other stakeholders, as appropriate in context, report identified AI system risk factors and/or circumstances that could result in impacts or harms:
  - ○ To organizational operations, including:
    - ■ Missions and functions
    - ■ Image and reputation, including:
      - Loss of trust and reluctance to use the system or service
      - Internal culture costs that impact morale or productivity
  - ○ To organizational assets, including legal compliance costs arising from problems created for individuals
- **For both internal and external stakeholders, as appropriate in context, report identified AI system risk factors and/or circumstances that could result in impacts or harms**:
  - ○ To other organizations
  - ○ To individuals, including impacts to health, safety, well-being, or fundamental rights
  - ○ To groups, including populations vulnerable to disproportionate adverse impacts or harms
  - ○ To the Nation or other societal impacts, including:
    - ■ Damage to or incapacitation of a critical infrastructure sector
    - ■ Economic and national security
    - ■ Impacts on democratic institutions and quality of life
    - ■ Environmental impacts
    - ■ **Additional identified factors that could lead to severe or catastrophic consequences for society**, such as:[52]
      - Potential for correlated robustness failures or other systemic risks across high-stakes application domains such as critical infrastructure or essential services
      - Potential for other systemic risks, which can be accumulated, accrued, correlated or compounded at societal scale, e.g.:
        - ○ Potential for correlated bias across a large fraction of a society's population
        - ○ Potential for many high-impact uses or misuses beyond an originally intended use case; e.g., if an AI system is a cutting-edge large language model, "foundation model" or another highly multi-purpose / general-purpose AI system

---

[52] This list closely follows the related list of factors in Section 3.2.2.1.2, and should be updated if the material in Section 3.2.2.1.2 undergoes revisions.

- Potential for large harms from misspecified goals
- Other identified factors affecting risks of high consequence / catastrophic and novel or "Black Swan" events

- For contexts where the organization is required to characterize an AI system according to another AI classification or risk communication framework (such as the OECD framework for the classification of AI systems, or frameworks for model cards, datasheets, reward reports, factsheets, or transparency notes), use AI system risk assessment outputs as part of preparation for AI classification/characterization reporting.[53] (Or if the AI system is already classified/characterized with another framework, use the AI classification/characterization information to inform risk assessment.)

If and when AI system incidents occur, report incident information, as appropriate in context:
- Report as many factors as feasible for the fields in the AIID CSET taxonomy, such as description of the incident, severity of harm, and harm type (AIID 2021b). See AIID (2021b) for the full list of fields, definitions and rating-level criteria, where available.

---

[53] Model cards (Mitchell et al. 2019) include a model's primary intended use, out-of-scope uses, and ethics issues (which can include risks and mitigations). Datasheets for datasets (Gebru et al. 2018) include datasets' recommended uses (as well as potential risks and mitigation). Reward reports (Gilbert et al. 2022) include objectives specification information (e.g., optimization goals and failure modes), and implementation limitations. Related industry approaches include Microsoft's Transparency Notes (see examples at Microsoft 2022b), IBM's FactSheets (Hind 2020) and Meta/Facebook's System Cards (Green et al. 2022). The OECD framework for AI system classification includes information on AI system contexts, data and input, AI model, and task and output (OECD 2022).

# 4. Recommendations for Roadmap for Later Versions of AI RMF or Supplementary Publications

The following ideas seem important to address as soon as appropriate in the AI RMF. However, developing appropriate guidance for these issues appears to require more research and development, which may not be feasible for the AI RMF 1.0. We recommend that NIST include these in an AI RMF roadmap for versions of the AI RMF or supplementary publications after AI RMF Version 1.0, if it is not feasible to include guidance for them in AI RMF Version 1.0.

## 4.1 An AI RMF Profile for Cutting-Edge, Increasingly Multi-Purpose or General-Purpose AI

We believe it would be valuable for NIST to provide at least one AI RMF Profile specifically oriented towards managing the broad context and associated risks of increasingly multi-purpose or general-purpose AI (including "foundation models").[54] These AI systems, such as BERT, CLIP,  GPT-3, DALL-E 2, and PaLM, can serve as multi-purpose AI platforms underpinning many end-use applications. These increasingly powerful, increasingly multi-purpose or general-purpose advanced AI systems are the focus of cutting-edge research. They also have several qualitatively distinct properties compared to the more common, narrower machine learning models, such as potential to be applied to many sectors at once, and emergent properties that can provide unexpected beneficial capabilities but also unexpected risks of adverse events. An AI RMF Profile for increasingly multi-purpose or general-purpose AI could address important underlying risks and early-development risks of such technologies in a way that does not rely on great certainty about each specific end-use application of the technology. For example, the Profile could provide Map or Measure function guidance, such as on assessing potential for catastrophic risks to society such as correlated robustness failures across multiple high-stakes application domains such as critical infrastructure. (For more on the capabilities and risks of increasingly multi-purpose and advanced AI, see, e.g., Bommasani et al. 2021 and Russell 2019.)

Guidance in the AI RMF Profile for increasingly multi-purpose or general-purpose AI could be based in part on examples of assessments and/or risk management controls already implemented by market leaders such as DeepMind and OpenAI. For example, OpenAI's 2019

---

[54] In this document, we intend the term "increasingly multi-purpose or general-purpose AI" to include a range of AI systems, from present day multi-purpose machine learning models and "foundation models" that typically require some domain-specific fine-tuning for application development, to more advanced future AI systems exhibiting greater or broader capabilities without domain-specific fine-tuning. We intend our usage of the terms "multi-purpose AI" and "general-purpose AI" to be compatible with usage of the term "general purpose AI" in the OECD classification framework and draft EU AI Act. The OECD classification framework states that GPT-3 is an example of "a general purpose AI system, meaning it can theoretically be used to deploy applications in any sector of the economy." (OECD 2022, p. 64) The EU AI Act Presidency Compromise Text draft states that general purpose AI systems are understood as AI systems "that are able to perform generally applicable functions such as image/speech recognition, audio/video generation, pattern detection, question answering, translation etc." (EU 2021b, Section 70a)

announcement of GPT-2 included enumeration of several categories of potential misuse cases (OpenAI 2019a), which apparently informed OpenAI's decisions on disallowed/unacceptable use-case categories of applications based on GPT-3 (OpenAI 2020b). DeepMind's 2021 announcement of their large language model Gopher and 2022 announcement of their multi-modal and multi-task "generalist agent" Gato also included consideration of potential misuse, safety risks and mitigation (Rae et al. 2021; Weidinger et al. 2021; Reed et al. 2022).

Creation of an AI RMF Profile specifically for increasingly multi-purpose or general-purpose AI could provide industry with valuable risk-management best practices addressing their unique issues. For example, the Profile could provide guidance on sharing of AI RMF responsibilities between researchers and upstream developers such as OpenAI and DeepMind that create cutting-edge increasingly multi-purpose AI systems and offer AI platforms/APIs based on those AI systems in a manner that allows many different end uses, and downstream developers that build upon the AI platforms for specific end-use applications using upstream provider-supplied information that may not be customized for their own application area. We also believe it would be appropriate to carry out more in-depth risk assessment with longer time horizons, at more points in the AI system life cycle, and to implement other more extensive risk-mitigation controls, etc. for increasingly multi-purpose AI systems than for AI with more limited capabilities. For example, it could be valuable for red teams to conduct more extensive interaction with AI systems to identify emergent properties of such systems, which are more likely with large-scale machine learning models, though it also may be more difficult or impossible to detect emergent hazardous capabilities or other characteristics of increasingly advanced AI (Hendrycks et al. 2021 p. 7).

Creation of an AI RMF Profile for increasingly multi-purpose AI would allow the AI RMF to provide appropriately targeted guidance for a small number of increasingly multi-purpose AI systems while minimizing costs for a large number of other, more narrow AI systems. We believe that most AI systems could be readily identified as one of the following:

A. One of a few large-scale, cutting-edge, increasingly multi-purpose or general-purpose AI system platforms (including "foundation models"). These AI systems would be the main focus of this Profile, with some corresponding costs for upstream developers but also corresponding risk-management benefits when employing guidance in this Profile.

B. A relatively narrow-purpose end-use application that builds on a multi-purpose AI system platform. Some aspects of these end-use application AI systems would be constructively addressed by a few parts of the guidance in this Profile. Costs to downstream developers would likely be minimal when employing guidance in this Profile.

C. One of many small-scale and/or stand-alone narrow-purpose systems, which are not within scope of this Profile. We do not expect developers or deployers of these common AI systems to use this Profile for those AI systems, and thus we do not expect their costs to be substantially affected by this Profile.

Although some other AI risk frameworks, such as the draft EU AI Act, seem highly focused on AI end-use application areas, we believe NIST has already created precedents for analogues to NIST AI RMF Profiles aimed at issues that cut across end-use application areas. For the Cybersecurity Framework, NIST (2021e) provides several example Profiles for industry sectors (e.g., position, navigation and timing services) that seem analogous to AI end use application categories. However, other NIST Cybersecurity Framework Profiles focus on critical issues (e.g., ransomware risk management) that extend beyond specific software application end-use categories.

In the following subsection, we go into more depth with illustrative examples of potential guidance in an AI RMF Profile for cutting-edge, increasingly multi-purpose or general-purpose AI and foundation models.

## 4.1.1 Illustrative Examples of "Map" Guidance in Profile for Cutting-Edge, Increasingly Multi-Purpose or General-Purpose AI

Regarding AI lifecycle and when to assess risks:

- For larger machine learning models, iterations are often slower than typical Agile sprints. For larger models, the pipeline is often to pretrain a model, analyze, customize, analyze, (customize differently, etc.), then deploy and monitor, then decommission. (Here we use "analyze" as a shorthand for probing, stress testing, red teaming, monitoring in simulated environments, etc.)
- For larger models, "Map" activities to identify risks should also happen after model training, not just before model training.

In addition to any originally intended use cases, identify other potential use cases and misuse cases, per the guidance in Section 3.1.2.1 of this document.
- (This is particularly important for increasingly multi-purpose and general-purpose AI, which can have large numbers of uses resulting from emergent capabilities.)
- OpenAI's 2019 announcement of GPT-2 included enumeration of several categories of potential misuse cases (OpenAI 2019a), which apparently informed OpenAI's decisions on disallowed/unacceptable use-case categories of applications based on GPT-3 (OpenAI 2020b).

As part of risk analysis:
- Assess potential for catastrophic risks to society such as correlated robustness failures across multiple high-stakes application domains such as critical infrastructure, per the guidance in Section 3.2.2.1 of this document. (This is particularly important for increasingly multi-purpose and general-purpose AI, which have potential for use across many domains; see, e.g., Bommasani et al. 2021 and Russell 2019.)
  - As part of staffing to identify potential catastrophic-risk scenarios for cutting-edge, increasingly multi-purpose or general-purpose AI, consider broadening the team to include social scientists and historians to provide additional perspective on structural risks that could emerge from interactions between an AI system and other societal-level systems. (Zwetsloot and Dafoe 2019)

○ After rating potential impacts using the scale in Section 3.2.2.1.2 of this document or an equivalent scale, consider also characterizing potential impacts using quantitative risk assessment (e.g., by estimating health and safety risks in terms of potential fatalities or quality-adjusted life years). This is an example of a more in-depth risk assessment approach that, despite its challenges and limitations, can illuminate additional dimensions of the risks (such as by identifying which scenarios could cause orders-of-magnitude larger impacts to public safety than others) and inform prioritization of risks.[55]

● In the context of cutting-edge, increasingly multi-purpose or general-purpose AI, consider treating the following AI system attributes as high-risk factors, indicating that corresponding targeted safety and control measures would be appropriate for those AI systems:

○ AI systems that could recursively improve their capabilities by modifying their algorithms or architectures through code generation (e.g., from OpenAI Codex or DeepMind AlphaCode), neural architecture search, etc.

■ Recursive improvement of AI system capabilities potentially could result in AI systems with unexpected emergent capabilities and safety-control failures. As the DeepMind paper on the software code-generation AI system AlphaCode stated, "Longer term, code generation could lead to advanced AI risks. Coding capabilities could lead to systems that can recursively write and improve themselves, rapidly leading to more and more advanced systems." (Li et al. 2022) For more, see, e.g., Russell (2019).

○ Adaptive models, which may be difficult to control in real time, e.g., in response to the coordinated manipulation attacks such as on the Microsoft Tay chatbot in 2016.

○ AI systems with ability to post entries in web forms, make HTTP POST requests, or use another outbound communication/influence channel. For related discussion, see, e.g., Nakano et al. (2021 p. 11), as well as general cybersecurity and software engineering resources on the principle of least privilege (for reasons to limit a system's privileges to the minimum necessary).

## 4.1.2 Illustrative Examples of "Measure" Guidance in Profile for Cutting-Edge, Increasingly Multi-Purpose or General-Purpose AI

Consider using red teams and adversarial testing as part of extensive interaction with state-of-the-art AI systems to identify emergent properties of such systems.

● Emergent properties are more likely with large-scale machine learning models than with smaller models, though it also may be more difficult or impossible to detect emergent hazardous capabilities or other characteristics of increasingly advanced AI (Hendrycks et al. 2021 p. 7).
- Consider automated generation of test cases as part of red team analyses. See, e.g., DeepMind's use of a language model for testing a version of the large language model Gopher (Perez et al. 2022).

---

[55] For brief discussion of quantitative risk assessment and approaches to refining risk assessments to inform prioritization, see, e.g., Ch. 2 and Appendix J of NIST SP 800-30. For additional discussion of challenges and limitations of quantitative risk assessment, including for expert-judgment and modeling methods often used in assessing risks of high-consequence, rare or novel events, see, e.g., Morgan and Henrion (1990) and Morgan (2017).

## 4.1.3 Illustrative Examples of "Manage" Guidance in Profile for Cutting-Edge, Increasingly Multi-Purpose or General-Purpose AI

Regarding pre-design and planning:
- Consider planning on deployment with gradual, phased releases, and/or structured access through an API or other mechanisms, with efforts to detect and respond to misuse or problematic anomalies. OpenAI has used a staged-release approach to roll-outs of large language models such as GPT-2, as well as a structured-access approach through an API for GPT-3, partly to minimize risks of misuse (OpenAI 2019d, Shevlane 2022). Meta AI is only providing full access to the large language model OPT-175B to researchers in academia, government, civil society, academia, industry research laboratories, and only for noncommercial research (Zhang et al 2022).
- Consider setting policies on disallowed/unacceptable use-case categories based in part on identified potential high-stakes misuse cases.
    - OpenAI's 2019 announcement of GPT-2 included listing several categories of potential misuse cases (OpenAI 2019a), which apparently informed OpenAI's decisions on disallowed/unacceptable use-case categories of applications based on GPT-3 (OpenAI 2020b). OpenAI also recommends publishing usage guidance as a best practice for large language models (OpenAI 2022).
- Consider setting policies on broader unacceptable-risk thresholds to prevent risks with substantial probability of inadequately-mitigated catastrophic outcomes.
    - For example, consider setting unacceptable-risk thresholds such that your organization would not develop or deploy cutting-edge AI agent systems with sufficient capabilities (such as robotics motor control, or sufficient generalist capabilities to learn robotics motor control) to cause physical harms, and with substantial chance of objectives mis-specification that currently cannot be adequately prevented or detected (such as deceptive alignment of advanced machine learning systems resulting from reinforcement learning or other training processes; see, e.g., Hubinger et al. 2019).
- If model training requires obtaining data sets, consider using only trusted training data instead of uncurated scrapes from the Web, to reduce vulnerability to backdoor and data poisoning attacks.
    - While data poisoning can be an issue for any machine learning model, this may be particularly challenging for training cutting-edge large models; often training of the largest new models has relied heavily on large-scale, uncurated Internet-scrape datasets (Bommasani et al. 2021 p. 106).
- Consider installing off-switch or circuit-breaker mechanisms for data centers before model deployment, or even before model training, to provide options for immediately shutting down data-center machines in case models exhibit safety-threatening behaviors as emergent properties.

- ○ Immediate-shutoff "kill switches" are a common safety feature in robots whose behaviors can result in physical harm.[56] These also may be appropriate as part of preparations for development and deployment of cutting-edge machine learning models with potentially emergent capabilities. Particular approaches to "safe interruptibility" may be needed to prevent advanced machine learning systems from circumventing an off-switch (see, e.g., Orseau and Armstrong 2016, Hadfield-Menell et al. 2016).

Regarding design and development:

- Consider disallowing open-ended learning with live web access; instead consider measures such as disallowing access to web forms (Nakano et al. 2021), disallowing HTTP POST requests, etc.
- Consider increasing the amount of compute (computing power) spent training state-of-the-art machine learning models only incrementally (e.g., by not more than three times between each increment) as part of management of risks of emergent properties.
  - ○ Often it is difficult to predict what failure modes machine learning models will have, what their performance will be, or what capabilities they will have. Machine learning systems are self-organizing systems that learn many features without explicit instruction. Incremental scaling-up approaches provide more opportunities for red-team monitors to identify emergent properties at an early or partially-emergent stage, when responses to identified emergent properties may be more feasible and effective. (For related discussion of emergent properties see, e.g., Section 3 of Hendrycks et al. 2021, and Bommasani et al. 2021.)
- Consider testing machine learning models after each incremental increase of compute, data, or model size for model training. If a large incremental increase (e.g., three times or more compute, or two times or more data or model parameters)[57] was used in a particular model training increment compared to the previous model training increment, it will be particularly important for the new model to be heavily probed/monitored/stress tested using detailed analysis processes (including or extending standard cybersecurity red team methods) to identify emergent properties such as capabilities and failure modes.

Regarding test and evaluation:

- After training and before deployment, probe/monitor/stress cutting-edge machine learning models using detailed analysis processes (including or extending standard cybersecurity red team methods) to identify emergent properties such as capabilities and failure modes.

---

[56] This is also related to the US Department of Defense (DoD) AI ethics principle of "governability": "DoD AI systems should be designed and engineered to fulfill their intended function while possessing the ability to detect and avoid unintended harm or disruption, and for human or automated disengagement or deactivation of deployed systems that demonstrate unintended escalatory or other behavior." (DIB 2019)

[57] For more in-depth discussion of relationships between scaling of compute, data and model size, see, e.g., Section 3.4 of Hoffman et al. (2022).

- Check for backdoors/AI trojans during testing/evaluation, especially for models trained on untrusted data from public sources with susceptibility to data poisoning. Tools to consider using include TrojAI (Karra et al. 2020, NIST n.d.b).

To further improve reliability in design and development, test and evaluation, and in deployment:
- Consider approaches to design, testing and deployment so that AI systems possess the minimum necessary capabilities for high reliability operation and not more capabilities.
- Consider methods of implementing the cybersecurity principle of least privilege. For example, consider using or extending typical "deny by default" or whitelisting methods, to limit an AI system's privileges to the minimum necessary for: access to information; communication channels; and action space.

## 4.1.4 Illustrative Examples of "Govern" Guidance in Profile for Cutting-Edge, Increasingly Multi-Purpose or General-Purpose AI

- Researchers and upstream providers creating cutting-edge increasingly multi-purpose AI systems and offering AI platforms/APIs based on those AI systems in a manner that allows many different end uses should be responsible for AI RMF tasks for which they have substantially greater information and capability than others in the value chain, such as:
  - Assessing and mitigating early-stage development risks
  - Identifying and disallowing potential high-stakes misuses
  - Making necessary information available to downstream developers and deployers building on "foundation model" AI platforms
    - Make as much information available on AI risk factors as reasonably possible to all audiences[58]
    - Provide additional information to downstream and end-use application developers and deployers as appropriate to meet their risk management needs
- Downstream developers and deployers building upon the AI platforms for specific end-use applications using upstream provider-supplied information that may not be customized for their own application area should be responsible for AI RMF tasks for which they have substantially greater information and capability than others in the value chain, such as:
  - Establishing specific AI RMF context for their intended end-use application(s)
  - Utilizing information provided by the upstream provider of an AI system platform, and requesting additional information as needed
- Downstream developers and deployers building upon cutting-edge platform AI systems should consider applying relevant parts of the guidance in this Profile for any substantial extensions of the underlying platform AI systems. Fine-tuned versions of the underlying platform systems often have capabilities that underlying platform systems do not.

---

[58] See Section 3.4.2.1 "Govern" Guidance for Reporting Information on AI Risk Factors and Incidents for draft guidance on providing stakeholders information on reasonably foreseeable risks without providing adversaries too much information.

# 5. Additional Ideas to Consider for Roadmap for Later Versions of AI RMF or Supplementary Publications

In this section, we list additional recommendations for items for NIST to consider adding to an AI RMF roadmap if they are not part of AI RMF 1.0. We believe these could have substantial value, but more research may need to be done before NIST can consider adding them to an AI RMF roadmap.[59]

## 5.1 Comprehensive Set of Governance Mechanisms or Controls to Help Organizations Mitigate Identified Risks

We expect it will be valuable for NIST to provide guidance for determining who should be responsible for implementing the Framework within each organization, as well as guidance regarding ongoing monitoring and evaluation mechanisms[60] that protect against evolving risks from continually learning AI systems, support for incident reporting, risk communication, complaint and redress mechanisms, independent auditing, and protection for whistleblowers, among other mechanisms. (NIST may provide guidance on some of these as part of AI RMF 1.0, under the "Govern" function. However, we would recommend an AI RMF Roadmap include development of a more comprehensive set of governance mechanisms or controls to mitigate identified risks, if they are not included in AI RMF version 1.0.)

## 5.2 Objectives Specification (i.e. Alignment of System Behavior with Designer Goals) Characterization and Measurement

AI system objectives specification (or alignment of system behavior with designer goals) can be regarded as a set of practices for the Safety characteristic in the AI RMF taxonomy.[61] DeepMind Safety Research (DSR 2018) includes "specification" as one of three key types of safety characteristics. (The other two are "robustness" and "assurance," which may map well to NIST's usage of those terms for trustworthiness aspects and procedures in the AI RMF.)[62] Specification represents the process of specifying AI system goals, objectives, or proxy metrics so that the system's behavior aligns with the designer's intentions. Specification problems occur when a system meets its literal goals, but also causes harms or has other behaviors that the designer did not anticipate or intend. Rudner and Toner (2021d) provide brief examples, such as social-media content recommendation machine-learning algorithms that learn to optimize user-engagement metrics by serving users with extremist content or disinformation. Rudner and Toner (2021d, p. 10) also suggest accounting for worst-case scenarios, and considering the following questions for an AI system, as part of identifying mis-specification risks: "What objective has been specified for the system, and what kinds of perverse behavior could be incentivized by optimizing for that objective?" An active area of AI safety research aims to develop methods for aligning AI systems during model training, and for validation and verification of AI system objectives alignment (see, e.g., Ouyang et al. 2022, and Bai et al. 2022; for more on challenges and future directions see, e.g., Section 4 of Hubinger et al. 2019, Gabriel 2020, Section 4.9 of Bommasani et al. 2021, and Section 4 of Hendrycks et al. 2021). These methods will be increasingly important as AI systems grow in capability. As mentioned in Section 1 of the NIST AI RMF Concept Paper, "Tackling scenarios that can represent costly outcomes or catastrophic risks to society should consider: an emphasis on managing the aggregate risks from low probability, high consequence effects of AI systems, and the need to ensure the alignment of ever more powerful advanced AI systems" (NIST 2021a).

---

[59] As an intermediate alternative to NIST placing these issues on an AI RMF roadmap, we may work to develop and test draft guidance on these issues ourselves, with a focus on incorporating the resulting draft guidance into recommendations on a Profile for cutting-edge, increasingly general-purpose AI. For several of these issues, such as objectives specification and recursive improvement, a focus on cutting-edge, increasingly general-purpose AI could yield large societal risk-reduction benefits while limiting costs to a small number of upstream developers of cutting-edge, increasingly general-purpose AI platforms.

[60] For some AI systems, it may be appropriate for an organization to both monitor their own system and also monitor other similar AI systems for indicators of potential issues with their own systems. Monitoring mechanisms also could include paying bounties to white-hat hackers to identify risks or failures, as in cybersecurity.

[61] AI system objectives specification and alignment can also be regarded as another guiding principle for AI development, consistent with the OECD principle "traceability to human values" and EU principle "human agency and oversight".

## 5.3 Generality (i.e. Breadth of AI Applicability/Adaptability) Characterization and Measurement

NIST could consider "assessment of generality" (i.e., assessment of the breadth of AI applicability/adaptability[63]) as another important characteristic affecting trustworthiness of an AI system, or perhaps as a factor affecting one or more of the AI trustworthiness characteristics NIST has already outlined. If an AI has not undergone any assessment of its generality, that would suggest lower trustworthiness. If assessment indicates high generality of an AI, we expect it would be appropriate to conduct more in-depth risk assessment, more assessment of use cases beyond the originally intended use cases, longer time horizons in risk assessment, more continuing assessment, etc. as we outline in the previous section proposing an AI RMF Profile for cutting-edge, increasingly multi-purpose or general-purpose AI. (Ideally, a generality assessment process would be quick and low-cost for the majority of AI with low generality, while accurately identifying the smaller number of AI with high generality.) For discussion of AI generality as a basic concept, see, e.g., Bommasani et al. (2021). For research on how to assess generality, see, e.g., Hernández-Orallo (2019) and Martínez-Plumed and Hernández-Orallo (2020).[64]

---

[62] The Georgetown University Center for Security and Emerging Technology (CSET) made these points in their submission in response to the NIST AI RMF RFI (NIST 2021b, Comment 26).
[63] In this document we use "adaptability" in the same sense as Bommasani et al. (2021), meaning an AI system's ability to be applied or adapted to solve many different tasks.
[64] Procedures for assessment of generality also may be a straightforward extension of our recommendations for consideration of the range of potential uses of an AI system.

## 5.4 Recursive Improvement Potential Characterization and Measurement

It could be valuable to assess the degree to which cutting-edge AI systems could recursively improve their capabilities, e.g., by editing their own training algorithm code through code generation or using neural architecture search. For such systems, greater levels of safety and control measures could be appropriate. As previously mentioned, in the context of cutting-edge, increasingly multi-purpose or general-purpose AI, recursive improvement potentially could result in AI systems with unexpected emergent capabilities and safety-control failures. As the DeepMind paper on the software code-generation AI system AlphaCode stated, "Longer term, code generation could lead to advanced AI risks. Coding capabilities could lead to systems that can recursively write and improve themselves, rapidly leading to more and more advanced systems." (Li et al. 2022) For discussion of related issues, see, e.g., Russell (2019).

## 5.5 Other Measurement/Assessment Tools for Technical Specialists Testing Key Aspects of AI Safety, Reliability, Robustness, etc.

AI safety researchers are working on a number of other concepts and measurement tools, many of which aim to address challenges in AI safety, reliability, robustness, etc. that are expected to grow as AI systems become increasingly advanced and powerful. NIST should maintain awareness of the leading AI safety concepts and tools, and consider them as appropriate when updating AI RMF guidance. See, e.g, Amodei et al. (2016), Ray et al. (2019), OpenAI (2019b, 2019c), and Hendrycks et al. (2021). CSET briefs on AI safety provide summaries for broad audiences on some of the key issues (Rudner and Toner 2021a, 2021b, 2021c). Measurement of AI safety risk properties is an active area of research; see, e.g., the discussion and references provided for Direction 1 ("Measuring and forecasting risks") in the 2021 Open Philanthropy request for proposals for projects in AI alignment (Open Philanthrophy 2021, Steinhardt and Barnes 2021).

# 6. Conclusion

In this document, we provide draft elements of actionable guidance on catastrophic risks and related issues, intended to be easily incorporated by NIST into the AI RMF version 1.0, or to serve as a complementary AI risk management practices resource for users of the AI RMF or other AI risk management guidance and standards. We also provide recommendations on additional issues for NIST to address as part of a roadmap for later versions of the AI RMF or supplementary publications. In addition, we provide our methodology for development of our recommendations. Our draft guidance aims to complement NIST's more general AI RMF procedures for risk assessment, mitigation, etc. to constructively address catastrophic-risk factors and related issues that NIST may otherwise not address in the AI RMF.

We aim for this work to be a concrete contribution to the NIST AI RMF, as well as to AI risk management guidance and standards more broadly. We welcome feedback on this document, with the aim of improving and building on it in revisions of this document and in follow-on work.

# Appendices

## Appendix 1: Scope and Time Frame of Risk Assessments and Impact Assessments in Related Risk Management Publications

In the following, we provide notes on passages related to risk assessment scope and time frame, in several relevant risk management publications.

The NIST AI RMF RFI and available draft AI RMF documents indicate that NIST intends for the AI RMF to consider potential impacts to individuals, groups, organizations and society. This language appears to be broadly consistent with the Cybersecurity Framework (NIST 2018) and NIST Privacy Framework (NIST 2020a). However, NIST has not fully defined those terms or associated procedures for the AI RMF.

The NIST Privacy Framework says little about scope or other aspects of how to perform risk assessment, other than referring readers to Brooks et al. (2017), i.e. NIST IR 8062. NIST IR 8062 (p. 22) notes that scope of impact assessment generally focuses on impacts to individuals, but "agencies may be able to use various other costs as proxies to help account for individual impact. They include, but are not limited to: legal compliance costs arising from the problems created for individuals, mission failure costs such as reluctance to use the system or service, reputational costs leading to loss of trust, and internal culture costs which impact morale or mission productivity as employees assess their general mission to serve the public good against the problems individuals may experience. Agencies also can consider expanding repercussions for the federal government, economic and national security, and societal impacts on democratic institutions and quality of life."  NIST IR 8062 (p.13) also references NIST SP 800-30 on cybersecurity risk assessment and NIST SP 800-37 on the NIST RMF lifecycle. (NIST SP 800-37 does not specifically mention time frames of risk assessment, and we do not discuss it further in the following.)

The NIST Cybersecurity Framework does not seem to clearly indicate scope of risk assessment on its own, but for risk assessment ID.RA-1 "informative references" it maps to CIS CSC 4, COBIT 5, ISA 62443-2-1, ISO/IEC 27001, and NIST SP 800-53 RA-3 among other NIST 800-53 controls.

NIST SP 800-53 RA-3 "discussion" (p. 243) indicates scope includes impacts to "organizational operations and assets, individuals, other organizations, and the Nation". NIST SP 800-53 RA-3 "discussion" (p. 243) on lifecycle: "Organizations can conduct risk assessments … at any stage in the system development life cycle. Risk assessments can also be conducted at various steps in the Risk Management Framework, including preparation, categorization, control selection, control implementation, control assessment, authorization, and control monitoring. Risk assessment is an ongoing activity carried out throughout the system development life cycle." NIST SP 800-53 references SP 800-30 Guide for Conducting Risk Assessments.

NIST SP 800-30 on scope of impact assessment: "Organizations determine potential adverse impacts in terms of organizational operations (i.e., missions, functions, image, and reputation), organizational assets, individuals, other organizations, and the Nation." (p. 26) "Table H-2 provides representative examples of types of impacts (i.e., harm) that can be considered by organizations." (p. 27)  NIST SP 800-30 Table H-2 (p. H-2) outlines types of adverse impacts, including harms to operations, to assets, to individuals (including "Injury or loss of life"), to other organizations, and to the Nation (including "Damage to or incapacitation of a critical infrastructure sector")  NIST SP 800-30 Table H-3 (p. H-3) outlines impact assessment scale including the following impact categories and descriptions:

- "Very high" impact: "The threat event could be expected to have multiple severe or catastrophic adverse effects on organizational operations, organizational assets, individuals, other organizations, or the Nation."
- "High" impact: "The threat event could be expected to have a severe or catastrophic adverse effect on organizational operations, organizational assets, individuals, other organizations, or the Nation. A severe or catastrophic adverse effect means that, for example, the threat event might: (i) cause a severe degradation in or loss of mission capability to an extent and duration that the organization is not able to perform one or more of its primary functions; (ii) result in major damage to organizational assets; (iii) result in major financial loss; or (iv) result in severe or catastrophic harm to individuals involving loss of life or serious life-threatening injuries."

NIST SP 800-30 on time frame as part of scope of risk assessment: "Effectiveness Time Frame: Organizations determine how long the results of particular risk assessments can be used to legitimately inform risk-based decisions. The time frame is usually related to the purpose of the assessment. For example, a risk assessment to inform Tier 1 policy-related decisions needs to be relevant for an extended period of time since the governance process for policy changes can be time-consuming in many organizations. A risk assessment conducted to inform a Tier 3 decision on the use of a compensating security control for an information system may be relevant only until the next release of the information technology product providing the required security capability. Organizations determine the useful life of risk assessment results and under what conditions the current assessment results become ineffective or irrelevant. Risk monitoring can be used to help determine the effectiveness of time frames for risk assessments. In addition to risk assessment results, organizations also consider the currency/timeliness (i.e., latency or age) of all types of information/data used in assessing risk. This is of particular concern in information reuse and evaluating the validity of assessment results." (p. 25)

NIST SP 1270 on time frame of algorithmic impact assessment, including steps to identify and address potential impacts of biases: "A misstep with impact assessments is to only apply them once at the beginning of a long and iterative process in which goals and outcomes can change over time. To overcome the challenge of the point-in-time nature of impact assessments, impact assessments must be applied at some reasonable cadence when used with iterative and evolving AI systems." (p. 36)

The draft EU AI Act (EU 2021a) does not currently appear to provide detailed procedures for AI system risk assessment, but it does define several risk categories based on use-case types and prohibition of unacceptable practices. The risk categories are "(i) an unacceptable risk, (ii) a high risk, and (iii) low or minimal risk" (EU 2021a Section 5.2.2). For example, AI systems would fall in the unacceptable-risk category if their use would violate fundamental rights. Our recommendations for risk assessment approach does not necessarily assume that organizations use "unacceptable risk" as a risk assessment category per se; instead we assume that the organization has established criteria for unacceptable risks based on regulatory compliance guidance, "risk appetite", and/or other risk acceptance.

ISO 31000 on risk management states generally that risk management processes should define scope and should consider time-related factors, given the context including organization-internal and -external factors. (The ISO/IEC 23894 AI risk management standard and ISO/IEC 42001 AI management system standard had not yet been publicly released when we performed this review.)

# Appendix 2: Questions for Reviewers in Google Form

Following are the questions that we requested reviewers answer, using a Google Form. To submit the form, responses to multiple-choice questions were mandatory, and responses to other (free-text) questions were not mandatory.

1. Does the draft guidance seem clear and actionable enough to be usable in context of the NIST AI RMF? (Options: Yes; No; Don't know)
2. Of the following, which would most improve the draft guidance? (Options: Adding usage flowcharts or other diagrams; Adding illustrative "worked examples"; Something else)
3. What else would be most valuable for improving the draft guidance?
4. Does the draft guidance seem compatible with Enterprise Risk Management (ERM) frameworks typically used by businesses and agencies? (Options: Yes; No; Don't know)
5. Does the draft guidance seem compatible with relevant standards or regulations, e.g. from NIST, ISO/IEC, IEEE, or the EU AI Act? (Options: Yes; No; Don't know)
6. Does the draft guidance seem usable or compatible with each stage of an AI lifecycle, e.g. design, development, test and evaluation, etc.? (Options: Yes; No; Don't know)
7. Does the draft guidance provide meaningful, actionable, and testable (i.e. "measurable") indicators of AI system trustworthiness, or at least enable documentability of risk management processes? (Options: It provides meaningful, actionable, and testable indicators of AI system trustworthiness; It at least enables documentability of risk management processes; None of the above; Don't know)
8. Is there LOW downside risk from publishing the draft guidance? (Does the draft guidance seem UNLIKELY to be misinterpreted/misapplied by users or other stakeholders in ways that would be net-harmful? Does publishing this guidance have LOW information hazards? Is the draft guidance sufficiently future-proof to be applied to AI systems over the next 10 years?) (Options: Yes; No; Don't know)

9. Overall, does the draft guidance meet or exceed its stated objectives enough to be a "minimum viable product" as part of guidance for the NIST AI RMF? (Options: Yes; No; Don't know)

10. If the draft guidance lacks something critical, what does it need to fill that gap?

11. Please provide your email address (optional) for feedback tracking and follow-up discussions.

# Acknowledgments

This work was financially supported by funding from Open Philanthropy. We appreciate feedback we received on earlier versions of this work from Helen Toner, Richard Mallah, Kim Lucy, Alexandra Belias, Jeremy McHugh, Rosie Campbell, Haydn Belfield, Markus Anderljung, Jonas Schuett, Andrew Barbe, Peter Cihon, Seth Baum, McKenna Fitzgerald, Caroline Jeanmaire, Sawyer Bernath, Luke Muehlhauser, Jan Leike, Evan R. Murphy, Sam Hilton, Shaun Ee, Andrew Critch, Carlos Gutierrez, Jason Green-Lowe, and others. Any remaining errors are our own.